\documentclass{article}
\usepackage{spconf,amsmath,graphicx,mathtools}
\usepackage{fancyhdr, lastpage, amssymb, float}
\usepackage{cite}
\usepackage{booktabs}
\usepackage{arydshln}

\title{Extended pipeline for content-based Feature Engineering in music genre recognition}
\name{Tina Raissi $^{\dagger }$ \qquad Alessandro Tibo $^{\ddagger}$ \qquad Paolo Bientinesi \sthanks {Financial support from the Deutsche Forschungsgemeinschaft (DFG) through grant GSC 11 is gratefully acknowledged.}$^{\dagger}$}

\address{
	$^{\dagger}$ RWTH Aachen University, AICES, Schinkelstr. 2, 52062 Aachen, Germany \\ 
	$^{\ddagger}$ University of Florence, Department of Information Engineering, Via S. Marta 3, 50139 Firenze, Italy}

\begin{document}
	%
	\maketitle

	\begin{abstract}
		We present a feature engineering pipeline for the construction of musical signal characteristics, to be used for the design of a supervised model for musical genre identification.\ The key idea is to extend the traditional two-step process of extraction and classification with additive stand-alone phases which are no longer organized in a waterfall scheme.\ The whole system is realized by traversing backtrack arrows and cycles between various stages.\ In order to give a compact and effective representation of the features, the standard early temporal integration is combined with other selection and extraction phases: on the one hand, the selection of the most meaningful characteristics based on \textit {information gain}, and on the other hand, the inclusion of the nonlinear correlation between this subset of features, determined by an \textit{autoencoder}.\ The results of the experiments conducted on {\it GTZAN} dataset reveal a noticeable contribution of this methodology towards the model's performance in classification task.
	\end{abstract}
	\section{Introduction}
	\label{sec:intro}
	One of the current subjects of research in Computer Science and Engineering concerns the enhancement of machines with abilities which are related to the human perception of the environment.\ Since recently, the term machine hearing\ \cite{lyon2010machine} is used as an umbrella to unify all applications of speech, music and environment sounds processing under a general concept.\ The central goal of this field is to model the hearing apparatus and its internal functionality \cite{gerhard2003audio}.\ However, in the case of speech and music, it is not possible to achieve this goal without taking into account the representation of intrinsic characteristics of the sound, developed in relation to the relative learning task.\ In our case, the quality of the musical signal features is evaluated with respect to the task of genre classification, a well-established, and rather controversial topic of Musical Information Retrieval.\\ 
	A quick look-back at remarkable works \cite{sturm2012survey} shows that the signal representation is obtained by the extraction of physical and perceptual features, in time, frequency and cepstral domains.\ This solution carries out the transformation of the original input signal to a new feature space.\ A reduction which results to be problematic, because the conservation of the relevant traits of the former is not guaranteed. This loss of information is caused by the applied analytical and computational model, and at the same time, by the rigid order of different steps of construction of the whole system.\ We shifted for this reason our attention from the analytical approach to a methodological one, by presenting a new feature extraction pipeline.\\ For some of the stages, our pipeline builds on existing methods in its isolated stages.\ We expanded the content-based features by adding the bottleneck layer's features of an \textit{autoencoder}, which forces to learn a low dimensional representation of the data.\ Moreover, the selection of the most predictive attributes, based on \textit{information gain} criteria, is done by using a \textit{Random Forests} classifier trained on an intermediate level feature vector. The entire process is not actually designed for recovering the lost information during the reduction, but for enriching the resulting feature vector with additive knowledge, useful for the specific classification task.\ The mean accuracy of the classifier, trained with the
	output dataset of this pipeline, is improved from the 78\% to
	91\%.
	
	In the following sections, we first introduce the general structure of this pipeline and describe every single process; we then present an evaluation of effectiveness of the features, at both final and intermediate stages.

	\section{General Pipeline}
	\label{sec:pipeline}
	
	For the most part, the proposed approaches in the literature for the automatic music genre identification, take into consideration a two-step process of extraction and classification, performed in consecutive order. In our work, the intermediate stages of the extraction gain autonomy, and can be reached from later stages.\ In this section we give a concise description of the process.
	
\begin{figure}[htb]
		
		\begin{minipage}[b]{1.0\linewidth}
			\centering
			\centerline{\includegraphics[width=8cm]{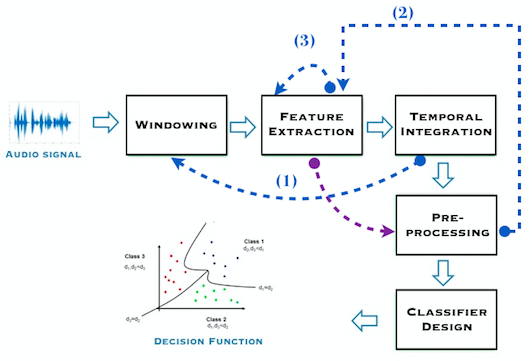}}
			\centerline{}\medskip
		\end{minipage}
		\caption{{The general pipeline for content-based feature engineering, with backtracking and forward arrows.}}
		\label{fig:pip}
	\end{figure}

	 As shown in Figure \ref{fig:pip}, the input of the process is a continuous audio stream.\ The extraction of information from the digitalized audio signal is generally carried out by algorithms operating on prefixed time window or consecutive blocks of frames.\ In order to obtain a vector $F \in \mathbb{R}^2$ relative to one feature, we should perform the stages windowing, feature extraction and temporal integration twice, by using the backtrack arrow (1).\ The first loop is responsible for the extraction of the \textit{short-time features}, while the second loop is concerned with the extraction of \textit{medium-time features} and the temporal integration over the whole signal length.\ In both cases we have an early temporal integration \cite{joder2009temporal} which uses the \textit{MeanVar} model \cite{meng2007temporal}.
	 
	 In Figure \ref{fig:pip}, the arrow (3) refers to the extraction of derivatives after the first windowing step.\ The feature vector extracted at this level can already be used for a classification task.\ In contrast to many existing approaches, we use this feature vector not for the final task, but for an intermediate classifier, built by random forests.\ The measure of the contribution of each attribute to the prediction of the target class is calculated by summing the information gain of each attribute, at every split.\ The attributes with positive contribution are then selected.\ In the pipeline, this step is indicated as preprocessing and is followed by another feature extraction step, connected via the backtrack arrow (2).\ This final extraction uses an \textit{autoencoder} which maps the vector of selected features to itself.\ The features selected from the bottleneck layer are not going to substitute the original dataset, but are added to it, as an additional information about nonlinear correlation between features.\ As a last step before the training phase, the dataset is normalized.

	\section{Feature extraction and taxonomy}
	\label{sec:feature}

	\subsection{Content-based features}
\begin{figure}[htb]		
			\begin{minipage}[b]{1.1\linewidth}
				\centering
				\centerline{\includegraphics[width=8cm]{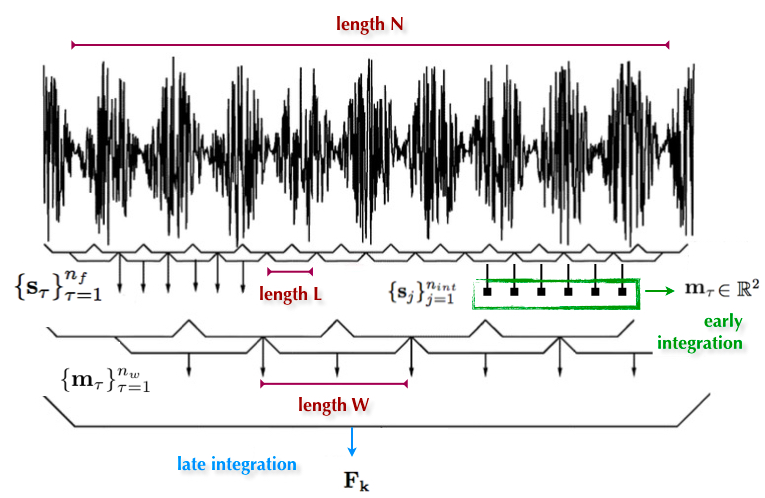}}
				\centerline{}\medskip
			\end{minipage}
				 \vspace{-2em}
			\caption{The temporal integration procedure relative to the extraction of one feature vector.}
			\label{fig:extract}
		\end{figure}

	We used three main categories of content-based features, based on a taxonomy presented in \cite{alias2016review}: Time Domain Physical, Frequency Domain Physical and Cepstral Domain Perceptual.\ Following the three basic requirements of musical characteristics introduced by [4], we are interested in obtaining a single value or a low-dimensional vector, from several feature observations, known also as early temporal integration method.\ We use the \textit{MeanVar} model, which calculates mean and standard deviation of observed values within a prefixed \textit{texture window}\cite{tzanetakis2002musical}.\ The early temporal integration procedure is illustrated in Figure \ref{fig:extract}: first, 14 features are extracted by using an \textit{analysis frame}\cite{tzanetakis2002musical} of 50 milliseconds with 50\% overlapping.\ Namely, Compactness, Energy, Entropy of Energy, Root Mean Square (RMS), Zero Crossing, 26 Mel Frequency Cepstral Coefficients (MFCC), Chroma vector corresponding to 12 semitones, the standard deviation of values of Chroma vector, 10 Linear Prediction Coefficients (LPC), Spectral Centroid, Spectral Flux, Spectral Rolloff, Spectral Spread and Spectral Variability.\ The computation of derivatives and a temporal integration step over both feature values and their derivatives follow.\ The Fraction of Low Energy Windows (FoLEW) is then determined, and a conclusive temporal integration step is performed.\ Finally, from the Beat Histogram the Beat Sum, Strongest Beat (SBeat), Strength of Strongest Beat (SSBeat), and their derivatives are extracted.\ The computation of all these features is detailed in \cite{lerch2012introduction}.

	\subsection{Early temporal Integration}
	
	As a first step, the signal is divided into $n_f$ analysis frames of length $L$.\ For every feature, a vector $\mathbf{s} \in \mathbb{R}^{n_f}$ of short-time values is extracted.\ In the second step, the mean and the standard deviation of $n_{int} = \frac{n_f}{n_w}$ values of the vector $\mathbf{s}$ corresponding to a texture window of length $W$ is calculated.\ Consequently, for every feature we have a matrix of medium-time values, having two rows derived from concatenation of $\mathbf{m}_{\tau} = [\mathit{\mu_{\tau}}, \mathit{\sigma_{\tau}}]^T, \forall \tau = 1, \cdots, n_w$.\ The feature vector $\mathbf{F}_k \in \mathbb{R}^2$ is the average value of every row of the medium-time matrix, corresponding to the $k$-th feature.
		
	\subsection{Bottleneck layer's features}
	
	 The linear version of \textit{Principal Component Analysis} (PCA) is a common learning method for analyzing and giving a low-dimensional representation of an input space\cite{jolliffe1986principal}.
	 
	  A generalization of PCA can be obtained by using an \textit{autoencoder}\cite{rummelhart1986parallel}, which is a pair of stacked \textit{neural networks}: an encoder $\mathbf{E}$ and a decoder $\mathbf{D}$.\ The encoder maps an input data $\mathbf{x} \in \mathbb{R}^n$ into a hidden vector $\mathbf{E(x)} = \mathbf{h} \in \mathbb{R}^d$ with typically $d \ll n$.\ The decoder maps $\mathbf{h}$ to an output vector $\mathbf{x}^{\prime} \in \mathbb{R}^n$.\ The \textit{autoencoder} is trained to copy the input $\mathbf{x}$ to $\mathbf{D(E(x))} = \mathbf{x^{\prime}}$ by using binary cross-entropy loss function:$$\mathcal{L}(\mathbf{x}, \mathbf{D(E(x))}) = -\big( \mathbf{x}\log(\mathbf{x'})+(\mathbf{1-x})\log(\mathbf{1-x'}) \big).$$

	If the data lies in a low dimensional manifold, the \textit{autoencoder} can actually learn such representation.\ Concerning music, we capture the evolution of the musical content in the signal by concatenating features referred to adjacent frames.\ The number of feature vectors that come from musically sound frames, is reasonably a subset of the whole feature space. 
	 	 	\begin{figure}[htb]
	 	 		\begin{minipage}[b]{1.1\linewidth}
	 	 			\centering
	 	 			\centerline{\includegraphics[width=7cm]{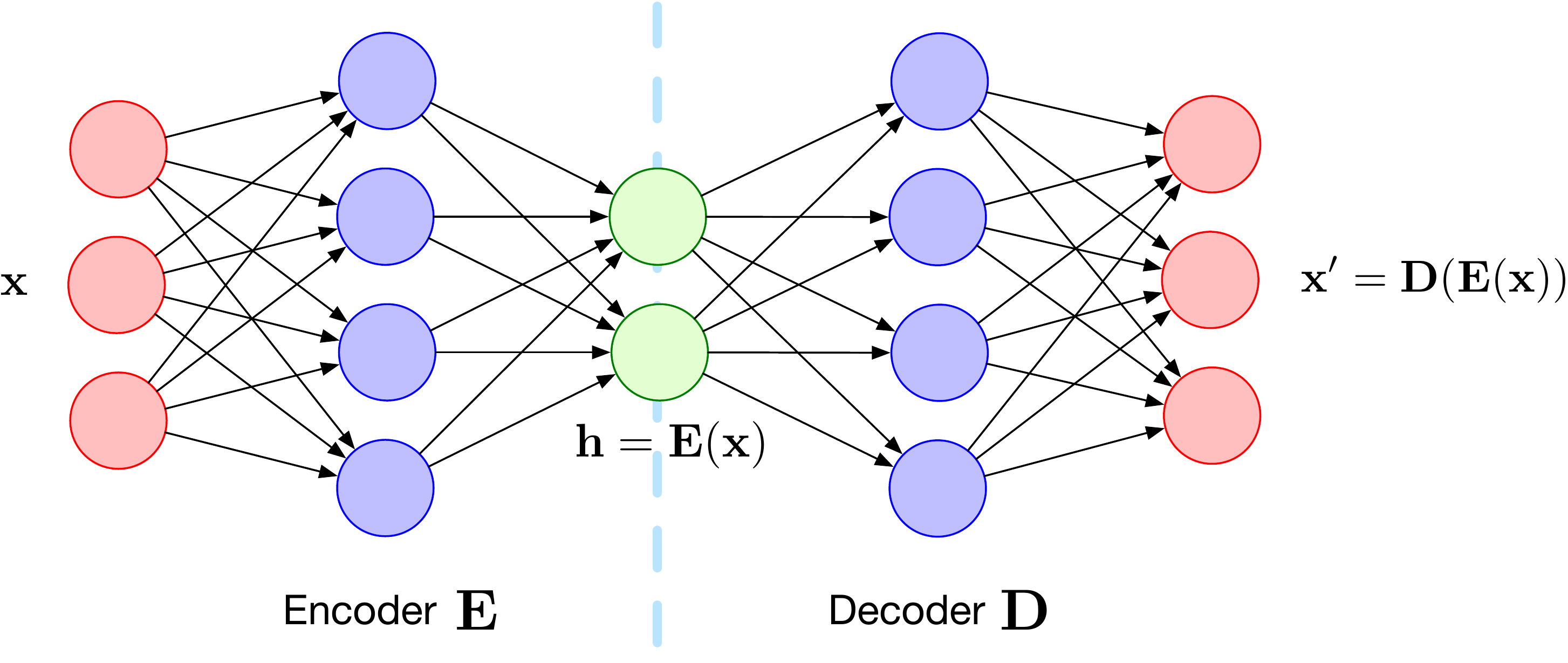}}
	 	 			\centerline{}\medskip
	 	 		\end{minipage}
		 	 	\vspace{-2em}
	 	 		\caption{The architecture of a symmetric autoencoder.}
	 	 		\label{fig:autoencoder}
	 	 	\end{figure}
	 	 	
The encoding and decoding functions in an \textit{autoencoder} has respectively $n_{e}$ and $n_{d}$ hidden layers, and in our case $n_{e} = n_{d}$.\ In addition to just mentioned $n_{e}+n_{d}$ hidden layers, we have also a middle layer known as \textit{bottleneck layer} which takes as input $\mathbf{E(x)}$.\ An example of an autoencoder is depicted in Figure \ref{fig:autoencoder}.

\subsection{Feature selection}
The notion of \textit{information gain} is relative to the average variation of information entropy due to current state's changes.\ Of particular interest is the application of this concept in \textit{decision trees}\cite{mitchell1997machine}.\ Generally speaking, the process of construction of a decision tree is based on the choice of the attribute with highest IG, on whose values the dataset of every node is split.\ In the feature selection process based on IG, every example is considered as a vector of values containing a specific information for predicting the class.\ We can construct a decision tree and determine the $IG_i$ of $i$-th attribute or feature by $\sum_{j=1}^{|nodes|}{IG_{i}}_j$, where ${IG_{i}}_j$ is the information gain obtained by splitting on $i$-th attribute at $j$-th node.\ The idea of ranking $IG_i$ values is equivalent to establishing an importance order of features, concerning their contribution to the prediction.

Given a set of training examples $\mathcal{S}$, with $\frac{|\mathcal{S}_{x_i = a}|}{|\mathcal{S}_j|}$ we denote the fraction of examples of the $i$-th attribute having value $a$ at $j$-th node.\ We define entropy as $$H(\mathcal{S}) =-\sum_{c \in classes}p_{c}(\mathcal{S})\log_2p_{c}(\mathcal{S}),$$
where $p_{c}(\mathcal{S})$ is the probability of a training example to belong to class $c$.\ Hence the information gain ${IG_{i}}_j$ obtained by splitting on attribute $x_i$ at node $j$ can be reformulated as 
$$IG(\mathcal{S}_j, x_i) = {H\left( \mathcal{S}_j\right)} - {\sum_{v\in values(x_i)}{\frac{\big|{{\mathcal{S}}_j}_{\left(x_i = v\right)}\big|}{|\mathcal{S}_j|}} H\left( {{\mathcal{S}}_j}_{\left(x_i = v\right)}\right)}.$$ This function measures the difference of entropies before and after splitting on the specific attribute.

\section{Classification}
\label{sec:class}
We consider two different classifiers as final models: the \textit{Support Vector Machines} with both radial basis and linear kernels, and \textit{Random Forests} for the feature selection step at intermediate stage of the pipeline.\ The choice of the latter, instead of a straightforward decision tree, derives from its well-known property of avoiding the overfitting problem.

	\section{Experiment}
	\label{sec:ex}
	The audio dataset is GTZAN \cite{tzanetakis2002musical}, consisting of 1000 audio tracks, each 30 seconds long and associated to one of ten genres: blues, classical, country, disco, hiphop, jazz, metal, pop, reggae and rock.\ Exactly 100 tracks are associated to every genre.\ All tracks were converted in 22,050 Hz mono 16-bit wav format.\ We ran 10 experiments splitting the dataset into 900 samples of training set and 100 samples of test set, preserving the percentage of samples for each class.\ Feature selection and hyperparameter optimization of SVM are done by 10-fold cross validation on the training set.\ For the extraction of content-based features, we used pyAudioAnalysis and jAudio \cite{giannakopoulos2015pyaudioanalysis, mckay2005jaudio}.\ The analysis frame for short-time features and the texture window for the medium-time features were set to 50 milliseconds and 1 second, respectively. In both cases we operated a 50\% of overlap.\ After the feature selection in the preprocessing step, we dropped all feature components which resulted in no information gain.\ Table \ref{tab:features} lists the selected features with their dimensionality corresponding to the experiment with the best result.\
	 
	 For the architecture of the autoencoder, summarized in Table \ref{tab:aeStructure}, we used three stacked hidden layers of size 60, 20, 60 respectively, with PReLU\cite{maas2013rectifier} activation, initialized using \emph{He normal initializer}\cite{he2015delving}.\ The first two hidden layers were followed by a Dropout\cite{srivastava2014dropout} layer with probability $0.2$.\ Finally an output layer of size 190 with sigmoid activation, initialized using \emph{He uniform initializer}, was stacked at the end of the network.\ The model was trained by minimizing the binary cross-entropy error loss.\ The autoencoder was fed with 900 training feature vectors of size 190, rescaled in $[0,1]$.\ We ran 100 epochs of the Adadelta\cite{zeiler2012adadelta} optimizer with learning rate 1.0, $\rho = 0.95$, $\epsilon = 1e-08$ without decay factor on minibatches of size 32.\ Even though we did not have sufficient resources for an automatic hyperparameter optimization, the proposed structure minimizes the loss function on the training set.
	 The final dataset consisted of content-based features of Table \ref{tab:features} augmented with bottleneck layer's features.\ Before the classification step the whole dataset was rescaled in $[0,1]$.\ The best final classifier, trained over this dataset, turned out to be the SVM with radial basis kernel.\ The hyperparameters of SVM resulted to be $\gamma = 2^{-6}$ and $C=4$.\ 
	 	 
	 	 		 	 \vspace{-1em}
	 	 		 \begin{table}[!h]

	 	 		 	\caption{ The content-based features and their dimensionality.\ For every feature, mean (M), standard deviation (SD), mean of derivatives ($\Delta$M) and standard deviation of derivatives ($\Delta$SD) are reported.\ The symbol "$\star$" indicates the elimination of feature components (see Section \ref{sec:feature} for the original dimensions) and "\text{\sffamily -}" means that no feature was extracted.\  }\label{tab:features}
	 	 		 	
	 	 		 	\begin{center}	 	 		 		
	 	 		 		\begin{tabular}{llllll}
	 	 		 			\toprule
	 	 		 			\textbf{Features} &  \textbf{M}  &  \textbf{SD}  & \textbf{$\Delta$M} & \textbf{$\Delta$SD} & \textbf{Total} \\
	 	 		 			\midrule
	 	 		 			Compactness          &   1 &    1 &         1 &          1 &     4 \\
	 	 		 			Energy               &   1 &    1 &         1 &          1 &     4 \\
	 	 		 			Entropy of Energy    &   1 &    1 &  \text{\sffamily -}  & \text{\sffamily -} &     2 \\
							FoLEW                &   1 &    1 &         1 &          1 &     4 \\
							RMS                  &   1 &    1 &         1 &          1 &     4 \\
							Zero Crossing        &   1 &    1 &         1 &          1 &     4 \\
							SBeat                &   1 &    1 &         1 &          1 &     4 \\
							SSBeat               &   1 &    1 &         1 &          1 &     4 \\
							Beat sum             &   1 &    1 &         1 &          1 &     4 \\ 
							MFCC                 &  22$^\star$ &   22$^\star$ &        15$^\star$ &         26 &    85 \\ 
	 	 		 			Chroma Vector        &  11$^\star$ &   11$^\star$ & \text{\sffamily -} &\text{\sffamily -} &    22 \\
	 	 		 			SD of Chroma         &   1 &    1 &\text{\sffamily -}& \text{\sffamily -}&     2 \\
	 	 		 			LPC                  &   9$^\star$ &    9$^\star$ &  0$^\star$ &         10 &    28 \\
	 	 		 			Spectral Centroid   &   1 &    1 &  0$^\star$  &          1 &     3 \\
	 	 		 			Spectral Flux        &   1 &    1 &         1 &          1 &     4 \\
	 	 		 			Spectral Rolloff     &   1 &    1 &         1 &          1 &     4 \\
	 	 		 			Spectral Spread      &   1 &    1 &         1 &          1 &     4 \\
	 	 		 			Spectral Variability &   1 &    1 &         1 &          1 &     4 \\	 	 		 			
	 	 		 			\bottomrule
	 	 		 		\end{tabular}
	 	 		 	\end{center}

	 	 		 \end{table}

	 	 		 	 \vspace{-2em} 
	 \begin{table}[!h]

	 	\caption{Autoencoder architecture.}\label{tab:aeStructure}
	 	\small
	 	\begin{center}
	 	\begin{tabular}{ll}
		 	\toprule
		 	\textbf{Layer} &  \textbf{\#Nodes} \\
		 	\midrule
	 		Dense, PReLU, Dropout(0.2)  &  60  \\
			Dense, PReLU, Dropout(0.2)   &   20\\
	 		Dense, PReLU  &  60  \\
	 			Dense, Sigmoid  & 190  \\ 
	 			\bottomrule
	 		\end{tabular}
	 	\end{center}
	 \end{table}	 
	 	 		 	 \vspace{-2em}
	\section{Results}
	\label{sec:res}
	In the Music Information Retrieval Evaluation eXchange (MIREX) 2012, the state of the art systems 
	for 10 genre categories achieved an accuracy of 50-80\% \cite{lerch2012introduction}.\ Table \ref{tab:ris} shows notable results in the literature.\ Regarding our approach, we measured the mean accuracy over ten experiments, at different stages of our pipeline: after the extraction of the content-based features, as the output of two cycles of windowing, feature extraction and early temporal integration steps, we obtained a mean accuracy of $78$\%.\ The preprocessing step of feature selection increased this result to the $86.30$\%.\ At the third and final stage, by performing an additional feature extraction step with autoencoder, we achieved $91$\%, thus surpassing the accuracy of all other approaches.

		\begin{table}[!h]
	
	\begin{center}
		\caption{Notable classification accuracies achieved in the literature for musical genre classification.}
			\vspace{0.5em}
		\label{tab:ris}
		\vspace{1ex}	
		\begin{tabular}{ll}
				 	\toprule
				 	\textbf{Reference} &  \textbf{Accuracy} \\
				 	\midrule
						Our approach  & $91.00$\% \\ 						
						Our approach, no bottleneck features & $86.30$\% \\
						Sturm et al.\cite{sturm2013music} & $83.00$\%  \\ 
						Bergestra et al.\cite{bergstra2006aggregate} & $82.50\%$ \\ 
						Li et al. \cite{li2003comparative} & $78.50$\%  \\
						Panagakis et al. \cite{panagakis2008music} & $78.20$\% \\
						Lidy et al. \cite{lidy2007mirex}& $76.80$\%  \\ 
						Benetos et al. \cite{benetos2008tensor}& $75.40$\%\\  
						Holzapfet et al.\cite{holzapfel2008musical}& $74.00$\%\\   
						Tzanetakis et al.\cite{tzanetakis2002musical}& $61.00$\% \\
						\bottomrule
			
		\end{tabular}
	\end{center}
\end{table}

\section{Conclusions}

We introduced a new feature engineering pipeline for the automatic identification of musical genre.\ In order to integrate the loss of information during the feature extraction, we reorganized the rigid waterfall scheme of the traditional two-step process of extraction and classification.\ We maintained continuity with respect to existing approaches in different fundamental aspects, like temporal integration method and feature taxonomy. Simultaneously we carried out a feature selection method based on information gain and extended the content-based features with bottleneck layer's features of an autoencoder.\ Neither of the two methods has ever been applied on the feature vectors extracted from the musical signal.\ Thank to the feature selection stage, our system achieved 86.30\% of accuracy.\ The extension of the preprocessed feature vectors with the bottleneck features had a further contribution of 4.7\% to the accuracy of the final model, which resulted to be 91\%.\

If we take into account the semantic ambiguity every genre recognition system must deal with, the results we obtained open new insights for the future works on definition and construction of genre and sub-genre identification systems, with an accuracy which can reach human performance.

	\vfill\pagebreak

\include{IEEEbib}
\bibliographystyle{IEEEbib}
\bibliography{strings,refs}

\end{document}